\documentclass[runningheads]{llncs}

\usepackage[T1]{fontenc}
\usepackage{graphicx}
\usepackage{amsmath}
\usepackage{hyperref}

\usepackage{subcaption}
\usepackage{placeins}
\usepackage{multirow}

\begin{document}

\title{A Study of Representational Properties of Unsupervised Anomaly Detection in Brain MRI}
\titlerunning{Representational properties of unsupervised anomaly detection}

\author{Ayantika Das\inst{1} \and
Arun Palla\inst{1} \and
Keerthi Ram\inst{2} \and\\
Mohanasankar Sivaprakasam\inst{1,2}}

\institute{Indian Institute of Technology, Madras, India \and
Healthcare Technology Innovation Centre, IIT Madras, India}

\maketitle

\begin{abstract}
Anomaly detection in MRI is of high clinical value in imaging and diagnosis. Unsupervised methods for anomaly detection provide interesting formulations based on reconstruction or latent embedding, offering a way to observe properties related to factorization. We study four existing modeling methods, and report our empirical observations using simple data science tools, to seek outcomes from the perspective of factorization as it would be most relevant to the task of unsupervised anomaly detection, considering the case of brain structural MRI. Our study indicates that anomaly detection algorithms that exhibit factorization related properties are well capacitated with delineatory capabilities to distinguish between normal and anomaly data. We have validated our observations in multiple anomaly and normal datasets. The code is available at \url{https://github.com/ayantikadas/Unsupervised_anomaly_detection}.
\end{abstract}

\section{Introduction}
Brain MRI is valuable in diagnosis, monitoring, and surgical intervention planning and guidance. Numerous signs and changes are visualized in brain MRI images, which could be pathological or normal variations~\cite{cenek_survey_2018}. MRI is versatile, and various sequences are available to observe visual features of several structural anomalies, including localized discrete (e.g., tumor, hemorrhage), diffuse (e.g., inflammation), regional and global (enlargement or contraction of structures, midline shift) changes. We focus on the detection of anomalies visible in the brain MRI, arising from patient related factors, explainable as geometric and photometric changes when related to normal appearance.

Normative brain atlases have been developed for quantitative neuroinformatics~\cite{nowinski_evolution_2021}
and are used in stereotaxic neurosurgery and electrode placement, along with diagnostic decision assistance in clinical applications like stroke management.
For comparative analysis, the methods assume that anomaly causes wider structural change than normal variations (we term this the compactness property of normal variations). 
Population averaged atlases and multiatlas methods \cite{mori_atlas-based_2013} are used to cover multiple anatomical phenotypes and widen the range of applicability. These methods use intensity based registration for computing a warp or map between the brain scan and the atlas, to standardize the data, and study differences.
The standardization can be construed as abstracting normal geometric variations through the warp and abstracting photometric variations to standardize across scanners and imaging parameters, leaving the apparent patient-related changes for comparison.

The resulting image can be ascribed to a simple additive model in 2D:
\begin{equation}
    image = g\_normalcy(sliceno) + anomaly + inhomogeneities
\end{equation}
Where $g\_normalcy$ is a generative process to compute a normative reference slice. We assume that anomaly effects an additive change over the generative process, co-existing in the same domain of intensities. For lesions that distort surrounding brain parenchyma (such as intra-axial lesions), the generative process includes local elastic deformation, following which the intensity variations continue to be additive. When using a population atlas such as MNI or Colin-27 \cite{holmes1998enhancement}, the generative process is trivially indexing of an atlas slice. Other data-driven methods (including ones described in the next section) built on normal data alone, can also be used for the generative process. The last component is residual uncorrelated inhomogeneities to model scanner and parameter variations left over beyond the standardization step.

Under this oversimplified model, the task of anomaly detection is reduced to one of decomposing the image domain; this however entails problems analogous to adaptive thresholding methods for detection -- bias to local contrast, affecting smaller and subtle anomalies, and also subtle global changes.

In this study, continuing with the premise of unsupervised comparative methods for modeling anomaly, we analyze some selected methods, restricting them to use only normal data and seek to elicit an understanding of their functioning from the standpoint of factorization.
% FIXME

\section{Approaches for modeling anomaly}
Supervised learning techniques, known to have achieved high performance in detection, localization, and delineation of pathologies in MRI~\cite{gudigar_brain_2020}, are specific to a few types of lesions like tumor and stroke~\cite{kaka_artificial_2021}, and biased by the available annotations at training time (which are laborious to obtain).

 Contemporary studies~\cite{atlason_segae_2019} explore unsupervised techniques, towards building disease-agnostic models for pathology detection and delineation~\cite{baur_autoencoders_2021}. Being trained only with normal data, test time pathological data (out-of-distribution) are expected to result in performance degradation in the task of reconstruction, offering a means of flagging anomaly. Auto-encoders are a well-known approach used in literature, formulated as jointly learning to encode image $x \in {\rm I\!R}^{m \times n}$(m and n are the height and width of the image) into an embedding $z \in {\rm I\!R}^{d}$, and learning to decode the embedding to reconstruct the image $\hat{x}$. Notationally, $ z = f_{\theta}(x) ; \hat{x} = g_{\phi}(z) $, where $f_{\theta}$ is the encoder and $g_{\phi}$ is the decoder. 

Compatible with unsupervised learning, the detection step is possible at the output (where a noticeable reconstruction error $\delta(\hat{x},x)$ is expected for the out-of-distribution anomalous test data) or at the embedding space (where embedding of the anomalous test data $z_a$ is expected to deviate from the normative embedding distribution $P(z_n)$). 

\subsection{Selected methods}
We select the following methods for modeling normal brain data: 
\begin{enumerate}
    \item \textbf{VAE} - VAE forms a baseline for the unsupervised anomaly detection setting we have considered.  VAE imposes $P(z_n)$ to be multivariate Gaussian with learnable mean and variance (weak i.i.d assumption). The underlying assumption is sufficiency of a simple Gaussian latent distribution for explaining normal data. VAE is implemented by including a Kullback-Leibler (KL) divergence regularization term along with a reconstruction term $\delta:|x-\hat{x}|_1$ in the network loss function and trained with data using routine optimization methods like stochastic gradient descent (SGD); 
    \item \textbf{FactorVAE}~\cite{kim_disentangling_2018} - FactorVAE also imposes $P(z_n)$ to be a multivariate Gaussian. Like VAE, the KL divergence and reconstruction term are included in the loss function, but it also includes an additional Total Correlation (TC) penalty for independence between variables in $z$. The independence in $z$ aids in the better encoding of meaningful normal features within the dimensions of $z$.

    \item \textbf{SSAE}~\cite{Baur_ssae_2020} - A Laplacian scale-space Auto-encoder, which declaratively performs image domain multiband factorization into high- and low-frequency content, processing the Laplacian pyramid band-wise through independent sparse convolutional Auto-encoders and forming the reconstruction image by combining the multiresolution outputs back into a Gaussian pyramid;
    \item \textbf{Glow}~\cite{kingma_glow_2018} - Generative Flow, learns a bijective mapping $f_\theta:X \rightarrow Z$ from the data space into  a multivariate standard Gaussian distribution.  The learning process is based on the \emph{change of variables} formulation.
    The reversible transformation follows due to the bijective nature, resulting from the architectural choice of layers. The selected methods are described in Sec.\ref{Appendix}.%The model is updated based on the objective function composed of a fixed latent distribution and the Jacobi of model.
\end{enumerate}

The latent space convention for the set of methods (VAE, FactorVAE, and SSAE) which are derivatives of the Auto-encoder family, is to have embeddings ($z \in {\rm I\!R}^{d}$) of much lower dimension as compared to the input image space. But, GLOW is architecturally bound to have an embedding space dimension equated to the total number of pixels in the image space ($d=m \times n$). The conventional usage of the low dimensional embeddings in the former set of methods is attributed to the fact that the latent space of the Auto-encoders must have limited information capacity to prevent them from learning trivial identity functions \cite{jing2020implicit}. Additionally, reconstructions of VAE suffer from image quality degradation if there are high levels of increment in the latent dimensions. The latent dimensions of VAE can be increased only up to a certain range, beyond which the generation process does not benefit from the introduced variability \cite{yeung2017tackling}.  Thus, we have carried out experimental evaluations adopting the conventional architectural formulation of the selected methods, allowing the inter-method dimensional variations in the latent space.

\subsection{Hierarchy of properties}
Our chosen methods exhibit some properties closely relatable to factorization at the latent embedding $z$ (via $f$ and loss terms), and the output $\hat{x}$ (via $g$ and loss terms). The specific properties of interest in increasing the order of capability for solving the task of unsupervised anomaly detection under consideration are discussed below:
\begin{enumerate}
    \item \textbf{Compact organization} - proximal representation of normal variations;
    \item \textbf{Separable representations} - distinctiveness within groups of variables in $z$ (or embedding subspaces) representing normalcy and anomaly, disentangling both the groups;
    \item \textbf{Controllable representations} - observe controlled changes in $\hat{x}$ via perturbing $z$;
    \item \textbf{Identifiable features} - visual features of image space captured in embedding subspaces.
\end{enumerate} 

As the properties are to be observed at $z$ and $\hat{x}$, we resort to the following intuitive ways of evidencing them:
\begin{itemize}
    \item Studying the two dimensional spectral embeddings\footnote{\href{http://surl.li/cirgr}{http://surl.li/cirgr}}; in order to analyze linkages between the latent vectors and comment upon their proximities. The spectral embeddings are computed using a non-linear dimensionality reduction technique which deploys spectral decomposition of $z$.
    \item Performing Local Outlier Factor (LOF)\footnote{\href{http://surl.li/cirgo}{http://surl.li/cirgo}} analysis on $z$ to examine the separability, in order to comment upon the distinctness of representations between the normal and anomalous sets. LOF gives a Local Outlier Detection (LOD) score, which is computed by estimating the local density deviation measured in a certain neighbourhood.
    \item Introducing variable perturbations on $z$ to investigate changes in $\hat{x}$ and infer upon the nature and extent of controllability in both the domains. 
\end{itemize}

\section{Experimental setup}
\label{Exp_setup}

We take MRI brain scans of only healthy individuals for the training phase and include both healthy and pathological scans for the testing phase. The healthy set consists of T2 weighted MRI scans from CamCAN \cite{taylor2017cambridge}, IXI \cite{IXI} and HCP \cite{van2012human} and the anomaly set consists of T2 weighted High Grade Glioma (HGG) data from BraTS 2017 \cite{menze2014multimodal}. The training and testing datasets follow the regimes as mentioned below:
\begin{itemize}
    \item Training set: We have considered a set of 100, 100, and 26 volumes from  IXI, CamCAN, and HCP datasets respectively; in order to impart better modeling capability in the presence of data from multiple sources.
    \item Test set: We have taken four sub-categories, in order to facilitate the investigation of the factorization properties under consideration:\\
    \begin{itemize}
        \item \textit{Normal}: A set of 100 volumes, 50 each from IXI and CamCAN;
        \item \textit{Noisy}: A simulated pathological dataset generated by adding random patches of Gaussian noise to the Normal set;
        \item \textit{BraTS}: A set 100 volumes from BraTS HGG dataset;
        \item \textit{Only Pathology}: Only the pathological extractions from the BraTS set, by masking out the healthy regions.
    \end{itemize}
\end{itemize}

 A sequence of standardization protocols is applied to minimize the variations within the images that are incurred during the acquisition process. We perform affine registration of all the images to Colin-27, which transforms the spacing of all axial slices to 0.5~mm with an in-plane resolution of 0.5 $\times$ 0.5~mm. We take corrective measures to remove inhomogeneities from the scans through N4 bias correction. The brain mask for the Colin-27 atlas is used to curate brain regions through the removal of the skull and other unrelated anatomical regions. Also, the intensity profile of the images is matched with the Colin-27 atlas by using histogram matching.
 
 For modeling the Auto-encoder based approaches that we have selected (VAE, FactorVAE, and SSAE), the generalised encoder-decoder architecture from \cite{baur_autoencoders_2021} is adopted. Although they share the same architecture, the representational learning of latent space is defined by the training protocol, which is different among them. VAE captures the latent space with a dimensionality value $d$ equal to 128. The training specifications for VAE include; training of the model for 80 epochs, usage of L1 loss for supervision of the reconstruction term, and the weightage to the KL Divergence term is adjusted using $\beta$-annealing methods till 10 epochs and then made consistent throughout \cite{heer2021ood}. FactorVAE shares a similar training strategy as VAE but with an additional Total Correlation (TC) penalty term. The weightage to the TC term is linearly annealed from 0 to 0.35 for 30 epochs and then set to a constant value throughout the training process of 60 epochs.
 % The SSAE model has a spatial latent representation of 16 X 8 X 8, three autoencoders are used to model the multi-scale laplacians, L1 loss is used to supervise the reconstruction loss at all the three level with equal weightage, 
The SSAE is modeled with three Auto-encoders operating at different levels in the Laplacian pyramid and is supervised with the L1 reconstruction loss at all three levels with equal weightage to each of them. The latent space is captured with dimensionality $d$ equal to 1024, 256, and 64 for all the three Auto-encoders respectively. The model is trained for 50 epochs.

For modeling Glow, the architecture from \cite{kingma_glow_2018} is adopted. The training is carried out till the data is transformed into a fixed latent representation, with its dimension $d$ equal  to 16,384. The objective function is based on the \emph{change of variables} formulation which consists of a Gaussian function and determinant of Jacobian of the model. The former term penalizes the deviation of the model's latent representation from the fixed Gaussian and the latter reflects a local change in the volume incurred by the model. Both the penalty terms are weighted equally throughout the training process of 50 epochs.
All the methods use Adam optimizer with an initial learning rate of $10^{-4}$ \cite{kingma2014adam}.

\begin{figure}[hbt!]
        \centering
        \begin{subfigure}{0.475\textwidth}
            \centering
            \includegraphics[width=1.1\textwidth, height=4.6cm]{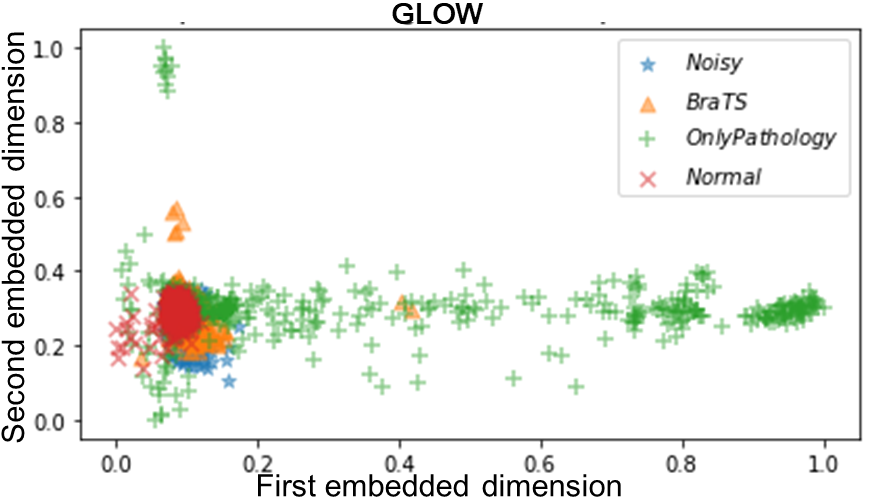}
            \caption[]%
            {{\small GLOW embeddings}} 
            \label{GLOW_embd}
        \end{subfigure}
        \hfill
        \begin{subfigure}{0.475\textwidth}  
            \centering 
            \includegraphics[width=1.1\textwidth, height=4.6cm]{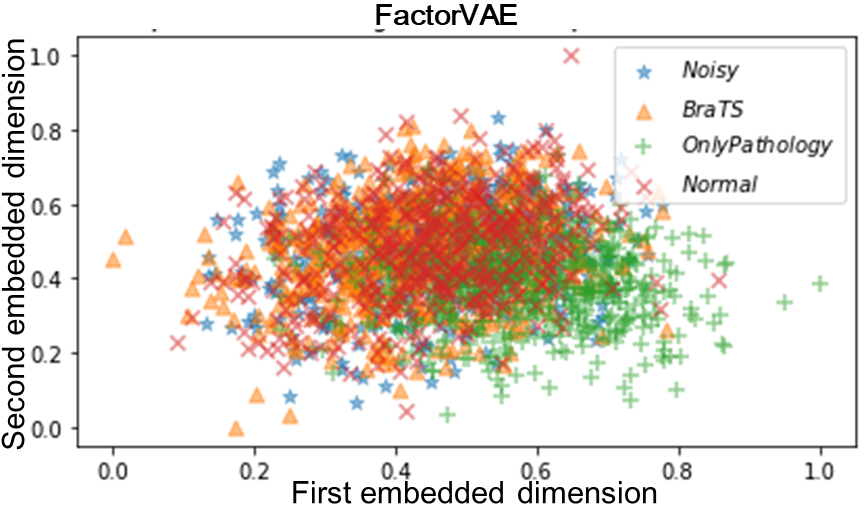}
            \caption[]%
            {{\small FactorVAE embeddings}}    
            \label{FactorVAE_embd}
        \end{subfigure}
        \vskip\baselineskip
        \begin{subfigure}{0.475\textwidth}   
            \centering 
            \includegraphics[width=1.1\textwidth, height=4.6cm]{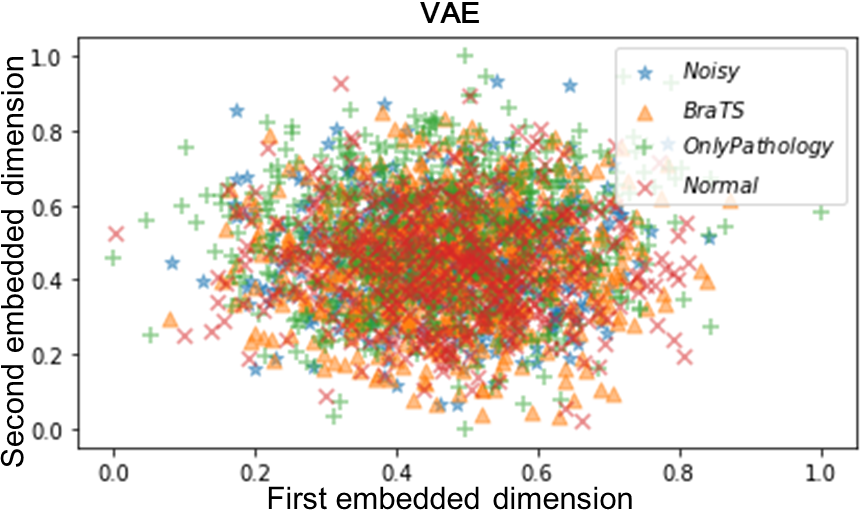}
            \caption[]%
            {{\small VAE embeddings}}    
            \label{VAE_embd}
        \end{subfigure}
        \hfill
        \begin{subfigure}{0.475\textwidth}   
            \centering 
            \includegraphics[width=1.1\textwidth, height=4.6cm]{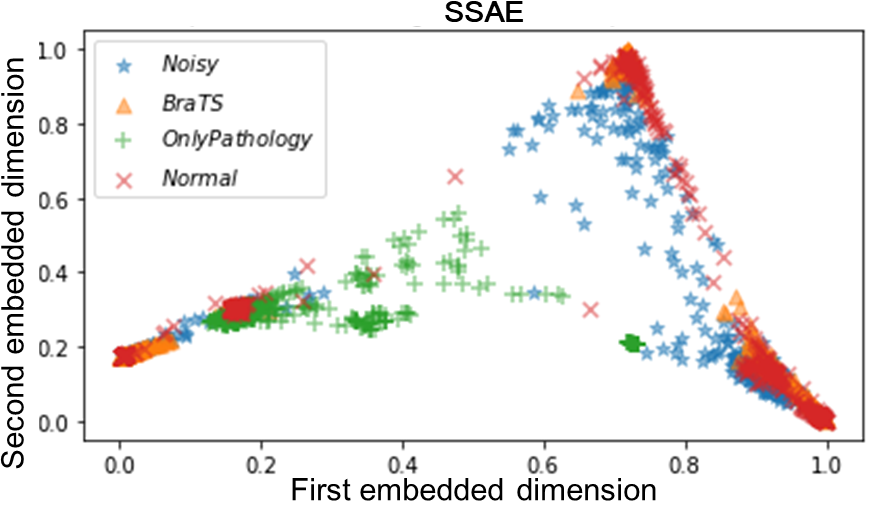}
            \caption[]%
            {{\small SSAE embeddings}}    
            \label{SSAE_embd}
        \end{subfigure}
        \caption[ ]
        {\small The two dimensional Spectral embeddings extracted from the high dimensional latent vectors of the Normal, BraTS, Noisy and Only Pathology test set categories as elaborated in Sec. \ref{Exp_setup} for all the four methods. The embedded space clearly depicts that the latent representations of the Normal set (in-distribution), from the (a) GLOW method, has a proximal and compact structure with distinctiveness as compared to the other out-of-distribution sets (BraTS + Only pathology + Noisy). The embedding space of the (b) FactorVAE and (c) VAE forms elliptical structures with less density. The distinctiveness property is completely missing from (c) VAE but the (b) FactorVAE has pushed away the only Pathology set to some extent. The embedding space from (d) SSAE neither adheres to any certain shape nor forms a compact structure.}
        \label{embd}
    \end{figure}
% \FloatBarrier
\section{Observations}
The observations pertaining to the four properties for the different test set categories; Normal, out-of-distribution (BraTS + Noisy + Only Pathology) are described below, highlighting comparative aspects between all the four methods.

\textbf{How compact and distinctive are the spectral embedding of the latent vectors?}

The Fig.\ref{embd} depicts the two dimensional spectral embeddings of the latent vectors. The organization of the latent vectors in the lower dimensional spectral embedding space is leveraged to carry out the analysis.
\begin{enumerate}
    \item \textit{GLOW:} From Fig.\ref{GLOW_embd} it is evident that the GLOW method is able to learn compact and proximal representations for the Normal set, which is surrounded by the representations of the other three out-of-distribution sets, with different degree of overlap. The degree of overlap in the Only Pathology set is very minimal with most of the samples pushed away from the Normal set in a less dense fashion. The BraTS and the Noisy set have relatively dense representations as compared to the Only Pathology set and their degree of overlap with the Normal set is high. \textit{Thus, GLOW is seen to  achieve a compact representation within the Normal set, with semantically relevant separation of the embeddings of the out-of-distribution set, disentangling meaningfully between the two sets.}
    \item \textit{VAE:} From Fig.\ref{VAE_embd}, it is seen that all the spectral embeddings of VAE are contained in an elliptical structure, which is much lower in density when compared to the GLOW embeddings. The Normal set does not uniquely hold the notion of proximal representations, rather the embeddings from all the sets are distributed in a similar manner. \textit{Thus, VAE presents a compact representation of embeddings by entangling both Normal and out-of-distribution sets within an elliptical structure, with no separation formed between any of the sets.}
    \item \textit{FactorVAE:} Similar to the VAE, the embeddings of the FactorVAE approximates an elliptical structural formulation as in Fig.\ref{FactorVAE_embd}. Compared to VAE, FactorVAE has relatively better proximity within the Normal set and distinctiveness between different sets; the deviation of the Only Pathology set from the concentrated region of the Normal set is visible. \textit{Thus, the notion of compact representations in FactorVAE is similar to VAE but it highlights better properties of separability.}
    \item \textit{SSAE:} The embeddings from SSAE are neither structurally bound nor proximal for any of the sets, as shown in Fig.\ref{SSAE_embd}. The samples from the Normal set have formed certain smaller groups in different regions within the embedding space. \textit{Thus, representations from SSAE are not compact for any of the sets, with multiple separations within the Normal set, making separations amongst sets less evident.     }
    
\end{enumerate}

\textbf{Which method shows delineation between Normal set and out-of-distribution set based on LOD score?}

\begin{figure}[hbt!]
        \centering
        \begin{subfigure}[b]{0.475\textwidth}
            \centering
            \includegraphics[width=1.1\textwidth, height=4.6cm]{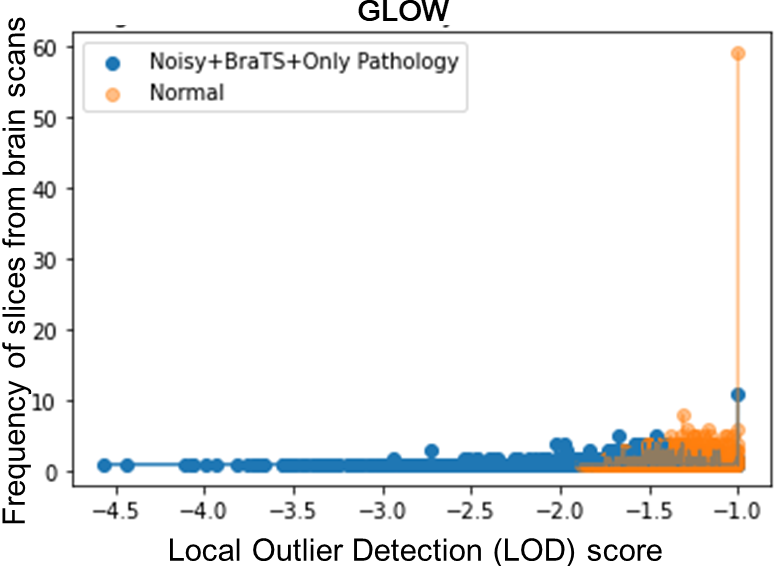}
            \caption[]%
            {{\small}}   %Histogram of LOD scores for GLOW
            \label{LOD_GLOW}
        \end{subfigure}
        \hfill
        \begin{subfigure}[b]{0.475\textwidth}  
            \centering 
            \includegraphics[width=1.1\textwidth, height=4.6cm]{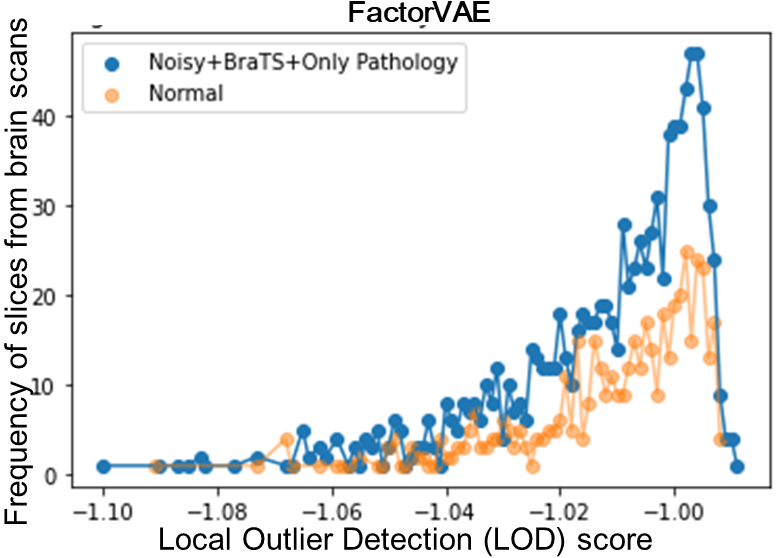}
            \caption[]%
            {{\small}}  % Histogram of LOD scores for FactorVAE  
            \label{LOD_FactorVAE}
        \end{subfigure}
        \vskip\baselineskip
        \begin{subfigure}[b]{0.475\textwidth}   
            \centering 
            \includegraphics[width=1.1\textwidth, height=4.6cm]{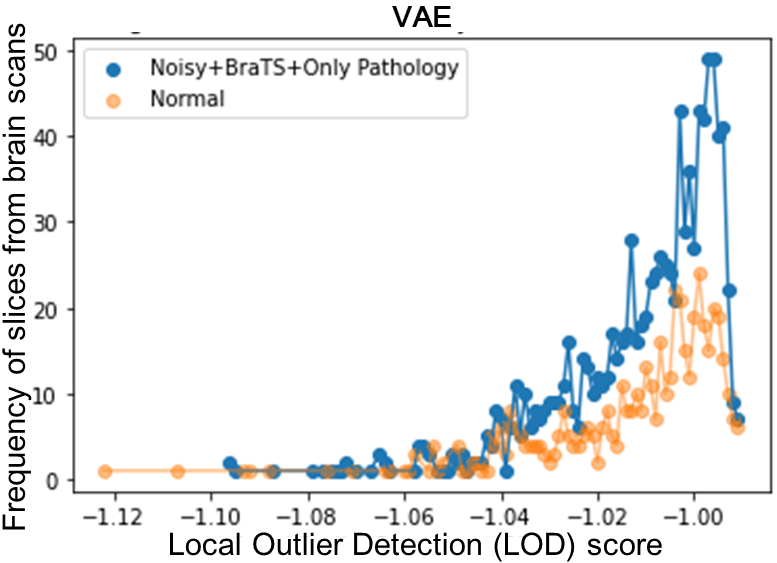}
            \caption[]%
            {{\small}}    % Histogram of LOD scores for VAE
            \label{LOD_VAE}
        \end{subfigure}
        \hfill
        \begin{subfigure}[b]{0.475\textwidth}   
            \centering 
            \includegraphics[width=1.1\textwidth, height=4.6cm]{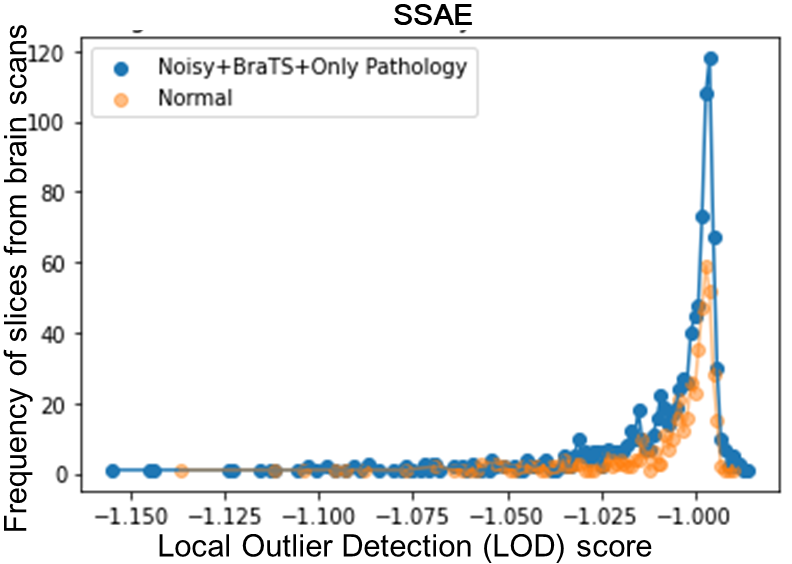}
            \caption[]%
            {{\small}}    %Histogram of LOD scores for VAE
            \label{LOD_SSAE}
        \end{subfigure}
        \caption[  ]
        {\small The histogram plot of the Local Outlier Detection (LOD) scores with the frequency of brain slices for the Normal set and out-of-distribution set (BraTS + Only pathology + Noisy). The LOD score is evaluated on the latent space components directly. The histogram plot from (a) GLOW highlights that the out-of-distribution set has a very heavy tail on the left hand side and the Normal set is mostly centered on the right side, showing delineatory properties. The histogram plot for (c) VAE, (b) FactorVAE, and (d) SSAE does not exhibit delineatory properties in the latent space.} 
        \label{LOD_}
    \end{figure}

The Fig.\ref{LOD_} represents the histogram of the LOD scores, calculated over the frequencies of brain slices for Normal and out-of-distribution sets. The histogram plot from \ref{LOD_GLOW} highlights that samples from the out-of-distribution set have a long tail on the left region of the plot and the Normal set is densely populated and contained within the right half. This behaviour \textit{showcases delineatory properties in GLOW}, despite having certain overlapping regions between the two distributions. The other histogram plots \ref{LOD_FactorVAE}, \ref{LOD_VAE} and \ref{LOD_SSAE} does not exhibit delineatory properties in the latent space.

\textbf{What is the nature of controllability in the reconstructed images ($\hat{x}$) due to different settings of perturbations in the latent vectors?}

\begin{figure}[tb]
    \centering
    \includegraphics[height=7cm]{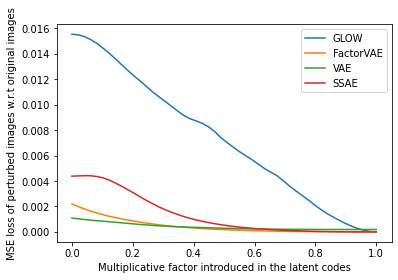}
    \caption{The dependency of latent space perturbations with change in the reconstructed image space, captured through Mean Squared error (MSE). Perturbations are in the form of $z*m$, where $z$ is the latent vector and $m$ is the multiplicative factor, with values ranging from 0 to 1. Effectively, an increasing order of perturbation in the latent space arises from decreasing values of $m$, implying $m=1$ has no change on $z$. MSE values are calculated between reconstructed images (due to varying values of $m$) and the original images. The GLOW has higher sensitivity to the perturbations followed by SSAE, FactorVAE, and VAE in decremental order of their sensitivity nature. }
    \label{Pertb_1}
\end{figure}

We have introduced scalar multiplicative factors ($m$) to perturb the latent space representations ($z*m$). The Fig.\ref{Pertb_1} highlights the dependency of the change in $\hat{x}$ due to the introduced perturbations in latent space.  All the adopted methods are sensitive and show monotonic increment to the perturbations, but the degree of sensitivity is method specific. The GLOW method exhibits the highest sensitivity to the perturbations as compared to the other methods. The VAE, FactorVAE, and SSAE show much staggered changes to the perturbations. \textit{Thus, GLOW allows highly sensitive and monotonic controllability to these perturbations as compared to other methods.  }

\begin{figure}[tb]
    \centering
    \includegraphics[ height=9cm]{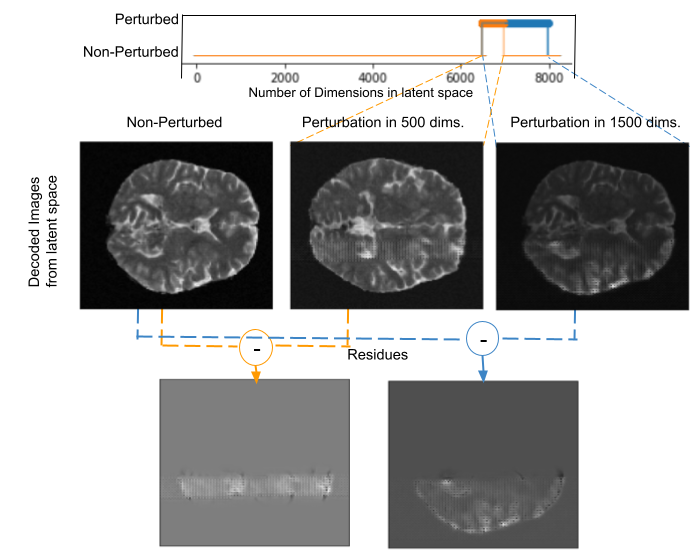}
    \caption{The top row pictorially depicts the introduced perturbations for the GLOW method with Case I and Case II, as in Table\ref{Table}. The second row highlights the reconstructed images (left to right) for non-perturbed $z$, and perturbed $z$ for 50 and 1500 dimensions respectively. The third row shows the residue (differences after subtracting) of the reconstructed images due to perturbations w.r.t the non-perturbed image. They are reflecting localized and proportional effects.}
    \label{GLOW_pertrb}
\end{figure}

We have experimented with another form of perturbation in the latent space, by corrupting a set of consecutive coordinates of the latent vectors, leaving all other coordinates undisturbed. The corruption is introduced in two folds Case I and Case II. For both the cases, out of all dimensions d in a latent vector, a certain set of consecutive coordinates p-q were changed and the remaining d-(p-q) were unchanged. Also for both the cases, p coordinate values are equal, but q coordinate values for Case II is greater than Case I, implying that perturbations in Case II are extended to a relatively larger span. Table\ref{Table} indicates the numerical values taken by p and q for all the methods in both the cases and the effect of these perturbations in $\hat{x}$. In both Case I and II, $\hat{x}$ from GLOW shows the effect of perturbation within certain consecutive rows, while the other methods show dispersed effects throughout the spatial extent of the images. For GLOW in Case II, when the perturbations get extended to consecutive coordinates of z, the $\hat{x}$ also transmits this to an extended set of consecutive rows. These sets of descriptive qualifiers from the Table\ref{Table} are pictorially depicted in Fig.\ref{GLOW_pertrb} for the GLOW method. The orange and blue notations signify Case I and Case II respectively.  \textit{Thus, GLOW shows localized and proportional controllability to these perturbations, while the other methods are only responsive to these changes without any specific nature. }

\textbf{How does latent space perturbations affect the identifiable visual features in the image space?}

From the Table\ref{Table}, we can comment that SSAE, FactorVAE, and VAE  disperse the effect of perturbations (within a few latent dimensions), over the entire image space (brain and background regions). But, \textit{GLOW is highlighting the effect of perturbations within the brain boundaries}, as shown in the residual images of the Fig.\ref{GLOW_pertrb}. Although regions within the brain boundaries are affected, but not necessarily different identifiable and meaningfully relevant compartments of the brain.

\begin{table}[]
\centering
\caption{Effect of perturbations in the latent space}
\label{Table}
\vspace{0.1cm}
\begin{tabular}{|c|c|cc|cc|}
\hline
\multirow{2}{*}{Methods} & \multirow{2}{*}{Dimensions of z (d)} & \multicolumn{2}{c|}{\begin{tabular}[c]{@{}c@{}}Perturbation defined\\  from p to q coordinates\\  of latent z\end{tabular}} & \multicolumn{2}{c|}{\begin{tabular}[c]{@{}c@{}}Location of the effect of\\  Perturbations  in \\ reconstructed image\end{tabular}}                                           \\ \cline{3-6} 
                         &                                  & \multicolumn{1}{c|}{Case I (p-q)}                                      & Case II (p-q)                                      & \multicolumn{1}{c|}{Case I}                                                                     & Case II                                                                    \\ \hline
GLOW                     & 16,384                           & \multicolumn{1}{c|}{6500-7000}                                         & 6500-8000                                          & \multicolumn{1}{c|}{\begin{tabular}[c]{@{}c@{}}Localized\\  within \\  Rows 75-95\end{tabular}} & \begin{tabular}[c]{@{}c@{}}Localized\\ within \\  Rows 75-125\end{tabular} \\ \hline
VAE                      & 128                              & \multicolumn{1}{c|}{50-90}                                             & 50-120                                             & \multicolumn{2}{c|}{\multirow{3}{*}{\begin{tabular}[c]{@{}c@{}}Dispersed over \\ the entire Image\end{tabular}}}                                                             \\ \cline{1-4}
FactorVAE                & 128                              & \multicolumn{1}{c|}{50-90}                                             & 50-120                                             & \multicolumn{2}{c|}{}                                                                                                                                                        \\ \cline{1-4}
SSAE                     & 1024                             & \multicolumn{1}{c|}{50-150}                                            & 50-600                                             & \multicolumn{2}{c|}{}                                                                                                                                                        \\ \hline
\end{tabular}
\end{table}


\section{Inferences and Discussion}
The relatability of the properties under consideration to the \textit{task of unsupervised anomaly detection} is discussed below, connecting to the observations from our methods and their reasoning.

\textit{Relating compact and separable representations to the task:} The separation between the representations of the Normal and out-of-distribution set is quite essential for modeling anomaly detection in the unsupervised setting we have considered. This holds from the fact that our unsupervised setting assumes representations of the out-of-distribution (test) set will be separated from the representations of the Normal (training) set, which will result in performance degradation in the former set indicating anomalies. Our observations have made us conclusive of the fact that separability within the representations is a consequence of the property of compact organization within the Normal set.
\begin{itemize}
    
    \item GLOW is the only method that is able to show compact organization within the Normal set and so it is able to disentangle and create separation between the sets. GLOW is able to show this property because it has learned to map the image space to a  high-dimensional standard Gaussian, whose dimensionality is equal to the image space. This has enabled it to capture attributes more meaningfully and semantically relevant to the image space. 
    \item VAE is entangling representations for Normal and out-of-distribution sets, enclosing representations from all these sets to an elliptical structure, since it has learned to map the image space to a much lower dimensional standard Gaussian. VAE needs to learn the reconstruction task to make the representations meaningful, while this task constrains the assignment of higher dimensions to $z$. This bottleneck entangles all the representations without any separation between them.
    \item FactorVAE also encloses representations within an elliptical structure since it shares a common objective with VAE but it has better separability since an additional term is introduced in the objective to ensure statistical independence amongst vectors in the latent space.
    \item SSAE is not capable of encoding compact and separable representation, since it learns a transformation for the Laplacian of the images at different scale spaces without any constraint in the formulation to the latent space.
\end{itemize}

\textit{Relating controllable representations to the task:} It is essential to observe controlled changes $\hat{x}$ while there are perturbations in $z$, for the task of anomaly detection that we have considered. In order to indicate the separation of the latent representations for Normal and out-of-distribution sets in $\hat{x}$, it is crucial for the anomaly detection task to transmit changes in $z$ to $\hat{x}$. The transmission of changes in $\hat{x}$ should be sensitive to the extent and location of perturbation in $z$ to localize anomalies.
\begin{itemize}
    \item GLOW method has the capability to control over $\hat{x}$, while there are perturbations in $z$, with high sensitivity and localizability as compared to other methods. These properties arise in GLOW due to the fact that it is architecturally designed to be bijective in nature, having a dimensionality of $z$ equal to $\hat{x}$.
\end{itemize}

\textit{Relating identifiable feature representations to the task:} The property of capturing the dimensions of $z$ with identifiable and visually meaningful compartments of the brain is hierarchically at the highest level of capability for our unsupervised anomaly detection task. This property can enable better capturing of the features for variations of normal brain information so as to reduce the chances of falsely detecting these normal variations as anomalies. The methods under consideration could not potentially encode information meaningfully about different compartments of the brain.

\section{Conclusion}
    We have performed an empirical study on the nature of latent representations of four existing models; GLOW, VAE, FactorVAE and SSAE by considering four properties; compact organization, separable representation, controllable features and identifiable visual features. We have visualized the latent representations in a reduced dimensional space, inferred on LOD score and introduced perturbations to compare the properties of the different methods. We observed that the GLOW method is able to show compact representations with meaningful separation and controllability. We have discussed explicitly how the properties relate to the task of unsupervised anomaly detection. We are able to conclude that GLOW is a potential candidate for the unsupervised anomaly detection since it exhibits better properties of factorization, which are closely related to the end task.
 
\FloatBarrier
\bibliographystyle{splncs04}
\bibliography{anomaly.bib}

\begin{thebibliography}{10}
\providecommand{\url}[1]{\texttt{#1}}
\providecommand{\urlprefix}{URL }
\providecommand{\doi}[1]{https://doi.org/#1}

\bibitem{IXI}
{IXI Datatset}. \url{http://brain-development.org/ixi-dataset/}

\bibitem{atlason_segae_2019}
Atlason, H.E., Love, A., Sigurdsson, S., Gudnason, V., Ellingsen, L.M.:
  {SegAE}: {Unsupervised} white matter lesion segmentation from brain {MRIs}
  using a {CNN} autoencoder. NeuroImage: Clinical  \textbf{24},  102085 (Jan
  2019). \doi{10.1016/j.nicl.2019.102085}

\bibitem{baur_autoencoders_2021}
Baur, C., Denner, S., Wiestler, B., Navab, N., Albarqouni, S.: Autoencoders for
  unsupervised anomaly segmentation in brain {MR} images: {A} comparative
  study. Medical Image Analysis  \textbf{69},  101952 (Apr 2021).
  \doi{10.1016/j.media.2020.101952}

\bibitem{Baur_ssae_2020}
Baur, C., Wiestler, B., Albarqouni, S., Navab, N.: Scale-space autoencoders for
  unsupervised anomaly segmentation in brain mri. In: Medical Image Computing
  and Computer Assisted Intervention – MICCAI 2020: 23rd International
  Conference, Lima, Peru, October 4–8, 2020, Proceedings, Part IV. p.
  552–561. Springer-Verlag, Berlin, Heidelberg (2020).
  \doi{10.1007/978-3-030-59719-1_54}

\bibitem{burgess2018understanding}
Burgess, C.P., Higgins, I., Pal, A., Matthey, L., Watters, N., Desjardins, G.,
  Lerchner, A.: Understanding disentangling in $\beta$ - {VAE}. arXiv preprint
  arXiv:1804.03599  (2018)

\bibitem{cenek_survey_2018}
Cenek, M., Hu, M., York, G., Dahl, S.: Survey of {Image} {Processing}
  {Techniques} for {Brain} {Pathology} {Diagnosis}: {Challenges} and
  {Opportunities}. Frontiers in Robotics and AI  \textbf{5} (2018)

\bibitem{gudigar_brain_2020}
Gudigar, A., Raghavendra, U., Hegde, A., Kalyani, M., Ciaccio, E.J.,
  Rajendra~Acharya, U.: Brain pathology identification using computer aided
  diagnostic tool: {A} systematic review. Computer Methods and Programs in
  Biomedicine  \textbf{187},  105205 (Apr 2020).
  \doi{10.1016/j.cmpb.2019.105205}

\bibitem{heer2021ood}
Heer, M., Postels, J., Chen, X., Konukoglu, E., Albarqouni, S.: The ood blind
  spot of unsupervised anomaly detection. In: Medical Imaging with Deep
  Learning. pp. 286--300. PMLR (2021)

\bibitem{holmes1998enhancement}
Holmes, C.J., Hoge, R., Collins, L., Woods, R., Toga, A.W., Evans, A.C.:
  Enhancement of mr images using registration for signal averaging. Journal of
  computer assisted tomography  \textbf{22}(2),  324--333 (1998)

\bibitem{jing2020implicit}
Jing, L., Zbontar, J., et~al.: Implicit rank-minimizing autoencoder. Advances
  in Neural Information Processing Systems  \textbf{33},  14736--14746 (2020)

\bibitem{kaka_artificial_2021}
Kaka, H., Zhang, E., Khan, N.: Artificial {Intelligence} and {Deep} {Learning}
  in {Neuroradiology}: {Exploring} the {New} {Frontier}. Can Assoc Radiol J
  \textbf{72}(1),  35--44 (Feb 2021). \doi{10.1177/0846537120954293}

\bibitem{kim_disentangling_2018}
Kim, H., Mnih, A.: Disentangling by {Factorising}. In: Proceedings of the 35th
  {International} {Conference} on {Machine} {Learning}. pp. 2649--2658. PMLR
  (Jul 2018), iSSN: 2640-3498

\bibitem{kingma2014adam}
Kingma, D.P., Ba, J.: Adam: A method for stochastic optimization. arXiv
  preprint arXiv:1412.6980  (2014)

\bibitem{kingma_glow_2018}
Kingma, D.P., Dhariwal, P.: Glow: {Generative} {Flow} with {Invertible} 1x1
  {Convolutions}. In: Bengio, S., Wallach, H., Larochelle, H., Grauman, K.,
  Cesa-Bianchi, N., Garnett, R. (eds.) Advances in {Neural} {Information}
  {Processing} {Systems}. vol.~31. Curran Associates, Inc. (2018)

\bibitem{menze2014multimodal}
Menze, B.H., Jakab, A., Bauer, S., Kalpathy-Cramer, J., Farahani, K., Kirby,
  J., Burren, Y., Porz, N., Slotboom, J., Wiest, R., et~al.: The multimodal
  brain tumor image segmentation benchmark (brats). IEEE transactions on
  medical imaging  \textbf{34}(10),  1993--2024 (2014)

\bibitem{mori_atlas-based_2013}
Mori, S., Oishi, K., Faria, A.V., Miller, M.I.: Atlas-{Based}
  {Neuroinformatics} via {MRI}: {Harnessing} {Information} from {Past}
  {Clinical} {Cases} and {Quantitative} {Image} {Analysis} for {Patient}
  {Care}. Annu Rev Biomed Eng  \textbf{15},  71--92 (Jul 2013).
  \doi{10.1146/annurev-bioeng-071812-152335}

\bibitem{nowinski_evolution_2021}
Nowinski, W.L.: Evolution of {Human} {Brain} {Atlases} in {Terms} of {Content},
  {Applications}, {Functionality}, and {Availability}. Neuroinform
  \textbf{19}(1),  1--22 (Jan 2021). \doi{10.1007/s12021-020-09481-9}

\bibitem{taylor2017cambridge}
Taylor, J.R., Williams, N., Cusack, R., Auer, T., Shafto, M.A., Dixon, M.,
  Tyler, L.K., Henson, R.N., et~al.: The cambridge centre for ageing and
  neuroscience (cam-can) data repository: Structural and functional mri, meg,
  and cognitive data from a cross-sectional adult lifespan sample. neuroimage
  \textbf{144},  262--269 (2017)

\bibitem{van2012human}
Van~Essen, D.C., Ugurbil, K., Auerbach, E., Barch, D., Behrens, T.E., Bucholz,
  R., Chang, A., Chen, L., Corbetta, M., Curtiss, S.W., et~al.: The human
  connectome project: a data acquisition perspective. Neuroimage
  \textbf{62}(4),  2222--2231 (2012)

\bibitem{yeung2017tackling}
Yeung, S., Kannan, A., Dauphin, Y., Fei-Fei, L.: Tackling over-pruning in
  variational autoencoders. arXiv preprint arXiv:1706.03643  (2017)

\end{thebibliography}
\section{Appendix}
\label{Appendix}
\subsection{VAE}
VAE \cite{burgess2018understanding} enforces the learning of marginal likelihood of the training data(x) in a generative setting. This learning process has an objective of maximizing the lower bound of the likelihood, written as:           

\begin{equation}
    \log p_{\theta}(x|z) \geq  L_{VAE} =\sum_{i=1}^{N} [E_{q_{\phi}(z|x)} [\log p_{\theta}(x_i|z)] -  D_{KL}(q(z|x_i) || p(z))] 
    \label{Appendix_vae}
\end{equation}
The notations $\phi$, $\theta$ denote the parameterization of the encoder and decoder in the VAE, z is the latent representation, N is the number of training samples, $D_{KL}$( || ) is non-negative Kullback–Leibler(KL) divergence, p(z) and $q_{\phi}$(z|x) are the prior and posterior distributions respectively.

Although from the theoretical perspective, there seems to be no bound on the prior and posterior distributions, but to make the optimization of the objective function in \ref{Appendix_vae} tractable both the distributions are approximated to be Gaussians with diagonal covariance matrix. Specifically, the prior distribution is set to standard Gaussians $P(Z) = \mathcal{N}(0,I)$. The encoder in our VAE setup takes healthy brain image $x$ as input and then estimates the mean and diagonal covariance matrix of the posterior for the given input sample. These estimated parameters are used to sample the latent vector corresponding to the input healthy scan through “reparameterization trick”. The reparameterization technique is introduced to enable gradient calculation with respect to the parameters of the encoder, which is generally not feasible in this scenario. The sampled latent vector is passed through the decoder to generate the reconstructed image $\hat{x}$.

\subsection{FactorVAE}

 FactorVAE \cite{kim_disentangling_2018} closely shares modeling components and optimization objectives with VAE along with some additional constraints on the latent space. The trade-off between the two objective components in VAE does not allow the posterior to achieve factorial nature similar to the prior. This factorial nature would enhance the independence of the latent vectors and thus would add more interpretability. The objective for FactorVAE is given as follows:

\begin{equation}
    L_{FactorVAE} = \sum_{i=1}^{N} [E_{q(z|x)}[\log p(x|z)] - D_{KL}(q(z|x )||p(z))] - \gamma D_{KL}(q(z)||q’(z))
    \label{Appendix_factorvae}
\end{equation}
where, q(z) = $E_{p(x)}$[q(z|x)], $q’(z)$:=product of (q($z_j$ )) (j=1: d), d is dimensionality of z. The objective \ref{Appendix_factorvae} augments a new component called Total Correlation (TC) to the objective function of VAE, which minimizes the KL divergence between the marginal posterior q(z) and its factorial representation $q’(z)$, but these terms are generally not tractable. So, a discriminator (D) is used to estimate the density ratio that arises in the TC term. The discriminator basically classifies latent vectors coming from each class q(z)/$q’(z)$ while encouraging the vectors to be factored.

\subsection{GLOW}
GLOW \cite{kingma_glow_2018} learns an invertible nonlinear transformation between input distribution and independent latent distribution. The invertibility is facilitated by the fact that the layer is designed to have bijective properties. Such bijective layers are cascaded to increase the capacity/expressiveness of the model. This non-linear but invertible mapping, together with the change-of-variables, would lead to a tractable and closed-form solution of the marginal likelihood representing the healthy images as in: 
\begin{equation}
    \log p(x) = \log p(z) + \log |det(df_\theta(x)/dx)| 
\end{equation}

where, p(z) is the prior, $f_\theta$ is the encoder, det($df_\theta$(x)/dx) determinant of Jaccobian of the model.

 This likelihood is decomposed in terms of 1) a fixed latent distribution, which is a standard Gaussians $P(Z) = \mathcal{N}(0,I)$  and 2) Jacobian of the model, constituting the training process. Since, GLOW learns an invertible mapping, explicit training for the decoder is not required. The encoder itself will uniquely determine the decoder. Due to bijective mapping, each point in the input space is uniquely represented by a latent code. Similarly, distinct latent codes are mapped to non-overlapping input images.

\subsection{SSAE}
SSAE \cite{Baur_ssae_2020} is an Auto-encoder based method that models normalcy without any additional constraint in the latent space and forms the most widely used baselines for pathology detection in unsupervised settings. SSAE disentangles the high and low frequency of the input data by learning to compress and reconstruct the laplacian pyramid of healthy MRI brain scans. Our application requires laplacian pyramids at four levels (K=0,1,2,3). Images at each level k are denoted as:
\begin{equation}
    I_k = d(g_\sigma  (I_{k - 1})) \qquad \forall \ 0 < k \leq K,
\end{equation}
where $I_0$ is the input image x, $g_\sigma$ (·) is a gaussian kernel with variance $\sigma$, d(·) is a downsampling operator. The high-frequency residuals $H_k$ at each level k are denoted as:

\begin{equation}
    H_k = I_k - u(I_{k+1}) \qquad \forall \ 0 \leq k < K 
\end{equation}
where u(·) is an upsampling operator. An image x is completely represented by the low-resolution image $I_K$ after K downsamplings and the high frequency residuals $H_0$, ...$H_{K-1}$. the final reconstructed image is obtained via the following:
\begin{equation}
    \hat{x} =\sum_{k=0}^{K-1} [ u(I_{K-k}) + H_{K-1-k} ]
\end{equation}

\end{document}